\input{aipcheck}

\documentclass[
    ,final            
  ]
  {aipproc}

\layoutstyle{6x9}

\begin{document}

\title{Modelling the flare activity of Sgr A*}

\classification{95.30.Sf, 97.10.Gz, 98.35.Jk}
\keywords      {Sgr A*, Light curve, Hot spot, Black hole physics, Galactic Center, Accretion disc, Relativity effects}

\author{E. M. Howard}{
  address={Department of Physics and Astronomy, Macquarie University, Balaclava Road, North Ryde, NSW, 2109 Australia}
}

\begin{abstract}
 
Latest observational data provides evidence that the emissions from
Sgr A* originate from an accretion disc within ten gravitational radii
of the dynamical centre of Milky Way.
We investigate the physical processes 
responsible for the variable observed emissions from the compact radio source Sgr A*.
We study the evolution of the variable emission region and analyse light curves and time-resolved spectra of
emissions originated at the surface of the accretion disk, close to the event
horizon, near the marginally stable orbit of a Kerr black hole. 

\end{abstract}

\maketitle


\section{Introduction}

Recent studies of the inner few parsecs at the centre of
our Galaxy provide clear indications of a supermassive black hole, object
associated with the unusual, non-thermal, variable radio source Sgr A*. The highly compact
nature of the distributed mass and the intense gravitational field suggest
the presence of dark matter. 

Sgr A* shows the feature of so-called flares, short bursts of increased emission that last for about 20-100 min, suggesting that the emitting region is located very close to the black hole. 
We investigate light curves from an emitting, variable hot spot co-moving with the accretion disk. Using relativistic ray-tracing methods, we produce time-resolved images and spectra of the accretion emission region, as seen by an observer located at infinity. The disk model is supposed to be keplerian, geometrically thin and optically thick in Kerr geometry.\cite{misner_1973} 

There is evidence that the emissions from
Sgr A* originate from the accretion disc from within only a few gravitational radii
from the black hole centre. In order to understand the physics behind the observed
time-dependent processes, we model light curves and spectra of
emissions originated at the surface of the accretion disk, close to the event
horizon, near the innermost stable circular
orbit (ISCO) of a Kerr black hole. \cite{carter_1968}
As seen in Figure 1, the photon trajectories originated at the surface of the accretion disk, near the last stable orbit, are calculated. This allows us to explore different emission models and various characteristics of the black hole. 
The observed spectra depend on the radiation emitted from the accretion disc
and it is influenced by the strong gravity field on its way to infinity.\cite{connors_1980}
The flux from an accretion disk mostly originates from the innermost regions,
where relativistic effects are very important for our understanding
of the observed variability patterns.\cite{matt_1991}
Relativistic effects play a crucial role, especially in the
time-dependent behaviour and particularly for a maximally rotating Kerr black
hole.\cite{fabian00} 

We take into account all relativistic effects such as light bending, gravitational lensing, redshift and time dilation. The variations in the redshift are evident in the spectrum shown in Figure 6. Using relativistic ray-tracing techniques, we analyze the constraining
parameters and characteristics of the black hole and study the major
imprint of both special and general relativistic effects on the time dependent
observed phenomena. \cite{matt92} 

\begin{figure}

\includegraphics[width=12cm]{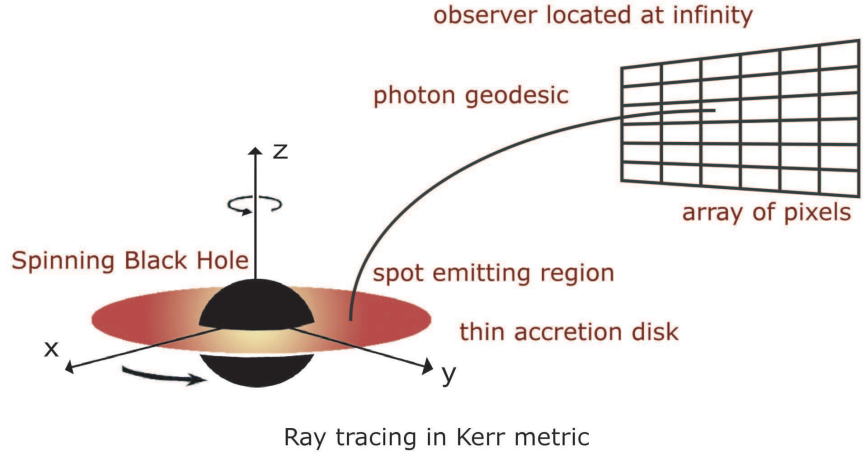}

\caption{Ray tracing in Kerr geometry}
\end{figure}

By integrating the photon geodesics between a position
inside a spot and a distant observer, we obtain time resolved spectra
for both orbiting and infalling spots co-moving with the accretion disk, in
a strong gravitational field. An example of a time dependent spectrum for a plunging spot can be seen in Figure 2.
We explore the evolution of the emission region, close to the event horizon and diagnose
light curves for various spectral emissivity
profiles, different viewing directions of the distant observer and different lo-
cations of the spot relative to the distant observer, the event horizon and the
center of the black hole. The emissivity is a function of both radial and 
and polar parameters.
The spot lightcurves may be interpreted as regions of enhanced emission in the local frame, near above the last stable orbit. Figure 3 is an example of a plunging trajectory of an infalling spot, the light curve of which is shown in Figure 4.

\section{Motivation}

Due to its proximity, SgrA* provides favorable circumstances 
to a better understanding of the processes responsible for the observed 
time-dependent phenomena.
Apparently, the emissions from Sgr A* are originated in  
radiative processes in keplerian motion, with a peak occuring within several
Schwarzschild radii ($r_S\equiv 2GM/c^2$) of the centre. \cite{walker_1970}

Latest theoretical models are trying to 
address the question of what the spectral line profiles or continuum
may tell us about the black hole properties, and
analyze the constraining parameters and characteristics of the accretion disc
close to the event horizon. \cite{ghisellini_1994} We take into account emissions from within 
the last stable orbit of a rotating black hole 
and from the region near above
the marginally stable orbit.\cite{beckwith04}

\begin{figure}

\includegraphics[width=10cm]{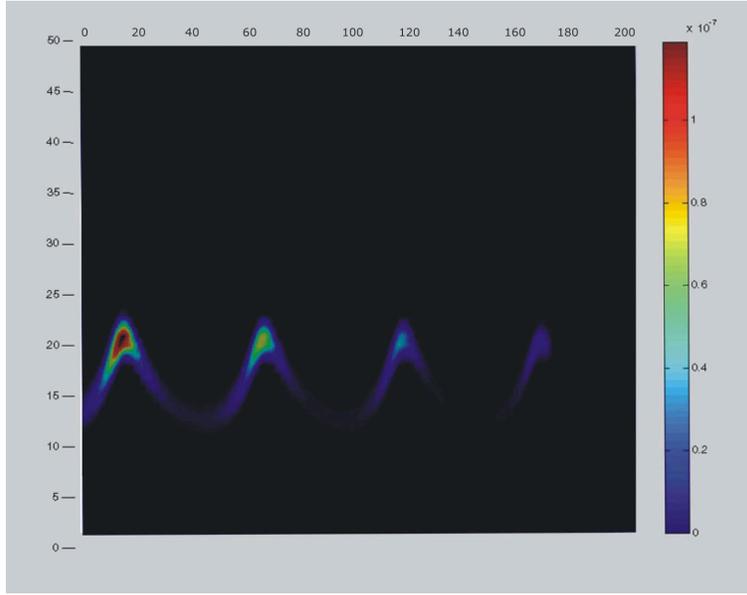}

\caption{Time-dependent spectrum of a spot falling from
the ISCO down towards the Kerr black-hole
event horizon. Energy is on the ordinate, time on
the abscissa}
\end{figure}

\section{Basic assumptions and objectives} 
We consider a complete system comprising a black hole, an accretion disc and a
co-rotating spot within the cold accretion disk. \cite{fanton97}
The gravitational field is described in terms of
Kerr metric\cite{nt73}, for a rotating black hole and particularly for a non-rotating black hole, in Schwarzschild metric.\cite{rauch94}. Figure 5 exemplifies the differences between the spectra obtained in Kerr and Schwarzschild metrics.
The co-rotating Keplerian accretion disc is geometrically thin and optically
thick, therefore we take into account only photons coming from the
equatorial plane directly to the sky plane. The spot is orbiting within
the disk near the rotating black hole.\cite{dovciak04}

A relativistic ray tracing code is used to
analyze how the considered effects affect the disk emissions 
originated in a co-moving (local) frame with the accretion disk, up to a observer
located at infinity and placed in the azimuthal direction $\varphi = \pi/2$. \cite{chandra92}
Relativistic effects alter the shape of the resulting spectra, 
more significantly at large inclination angles and play an important role
in the time-dependent emission processes, especially in the case of a 
maximally rotating Kerr black hole.\cite{kato98} The code is used with the standard
X-ray spectral fitting package, XSPEC (Arnaud 1996).\cite{arnaud96}

\begin{figure}

\includegraphics[width=11cm]{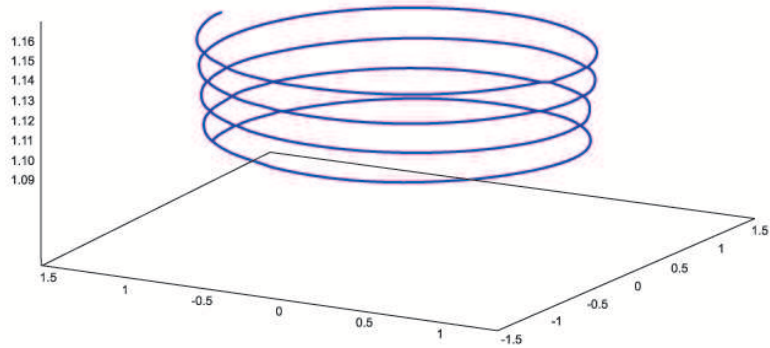}

\caption{Plunging trajectory of a photon into the event horizon of an extreme Black Hole}
\end{figure}

\section{Discussion}

We try to address several questions concerning the validity of the simple spot model and
the role of the relativistic effects in the observed processes.
We calculate time-dependent spectra of both cases of an orbiting and a free falling spot.
The code integrates the local emission in polar coordinates on the disc, enabling us to
handle non-stationary emission originated in a non-axisymmetric area of integration.\cite{dovciak04a}
In the axisymmetric case, the local emission is integrated in one
dimension, the radial coordinate of the disc.\cite{laor_1991}

\begin{figure}

\includegraphics[width=11cm]{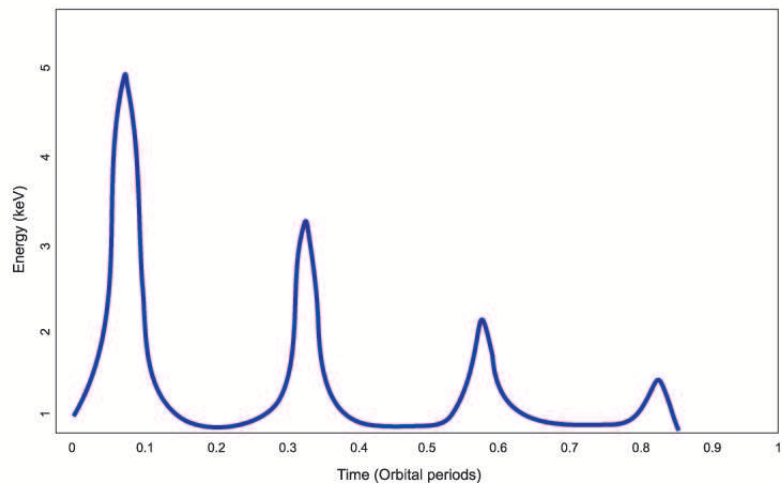}

\caption{Light curve of an in-falling spot near to the horizon of an extreme Kerr black hole.
with spin parameter a=0.998. Energy is on the ordinate, time on the abscissa.}
\end{figure}

We are able to:  
\begin{enumerate}
 \item gather information about the black-hole spin
 \item study plunging photon trajectories  starting at the marginally stable orbit down
towards the black hole event horizon. This provides us with information about 
the plunge region, the area between the Event Horizon and the last stable orbit.
 \item work with the emissivity as dependent on both of the polar coordinates within the equatorial plane
 \item deal with time variable spectra, therefore handle non-stationary cases, making possible the study of the time evolution of the emitting region. 
 \item study light curves for different inclination angles of the observer relative to the disk axis ($\theta_{\rm{}o}$) and analyze the dependence of the variability on the inclination.
 \item choose a non-axisymmetric emission area, therefore control the size and shape of the spot
 \item consider non-axisymmetric geometry of accretion flows
 \item analyze the general and special relativity effects that influence
photon paths along null geodesics towards an observer located at infinity \cite{krolik99}
 \item as the photon paths are integrated in Kerr ingoing
coordinates, this allows us to study the Kerr geometry and test the metric
\end{enumerate}

Emission starts within a localized area at the surface of the accretion disc.\cite{martocchia00} We consider
two cases, whether the spot moves with a Keplerian velocity along a stable circular orbit or, if close 
to the last stable orbit, it plunges into the black hole.\cite{connors_1977}
The intrinsic intensity at each point depends on the energy shift of the photons and it is time dependent.
\begin{figure}

\includegraphics[width=10cm]{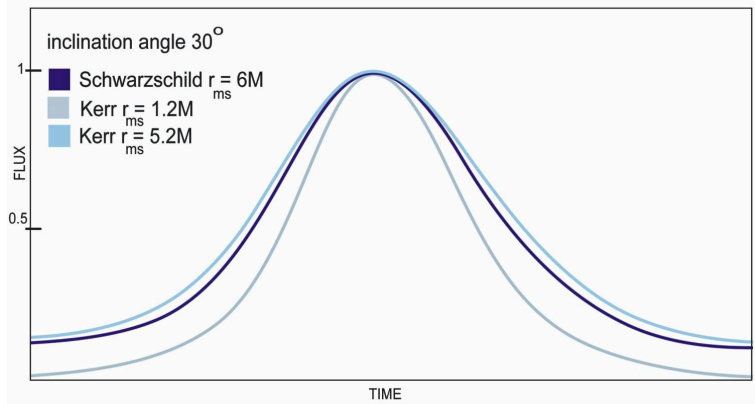}

\caption{Comparison between Kerr and Schwarzschild black hole light curves}
\end{figure}

Photons emerging from the spot 
would be affected by all relativistic imprints. \cite{chandrasekhar_1960}
The KYSPOT code by Dovciak et al.
takes all special and general relativistic effects into account
by using pre-calculated sets of transfer functions that map various properties of the emission region in
the accretion disk onto the sky plane (Cunningham 1975, 1976). Six functions are needed for the complete integration of the spectra of the emission region.

The transfer function, obtained by integration of the geodesic equation, correlates the flux in the local frame
comoving with the disk, to the flux as seen by the observer
located at infinity. \cite{schnittman04}
Spectral line profiles are obviously affected by relativistic smearing but due to the power-law character
of the relativistic imprints, the main shape of the primary continuum profile remains intact.\cite{reynolds03}

\begin{figure}

\includegraphics[width=10cm]{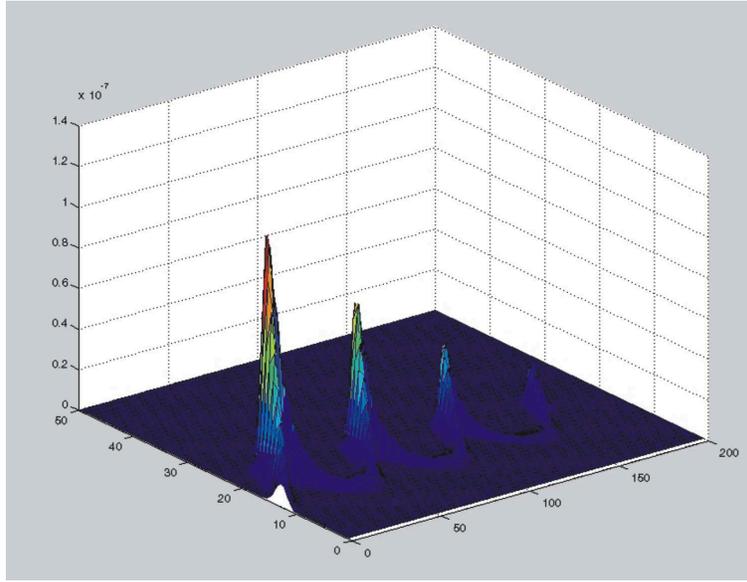}

\caption{Dynamic spectrogram of a plunging spot around a Kerr black hole. The profile includes variations in the redshift of the emitted light. }
\end{figure}
The functions are calculated for different values of
observer inclination angle and black hole horizons. 
The binary extensions contain information for different
radii and different values of $g$-factor, defined as 
the ratio between the photon energy observed at infinity
and the local photon energy as emitted from the disc. 
Because the emission from the disc is not stationary we take into account the
relative delay with which photons from different parts of the disc arrive to the distant observer.

We obtain time-dependent spectra for various
viewing directions of a distant observer based on different emissivity profiles and various angular momenta of the black hole.
The relativistic corrections on the local emission 
are parameterized by the black hole spin 
and the observer's inclination angle.\cite{gierlinski01}
The intrinsic emissivity is specified in the frame co-moving with the disc
medium and it is defined as a function of $r$, $\varphi$
and $t$ in the equatorial plane. When the geodesic integration is ended, after the transfer of photons to the distant observer is performed, Boyer-Lindquist coordinates will replace the initial Kerr ingoing coordinate system.
\cite{george_1991}
The observed spectra depend on the position of the spot with respect
to the disk normal. We obtain light curves of the variable emission region, for different positions and angles (azimuthal and polar) of the spot relatively to the distant observer, the event horizon and the center of the disk. We also consider different spin parameter values, various viewing angles and different sizes of the emitting spot.



\bibliographystyle{aipproc}   

\bibliography{sample}

\IfFileExists{\jobname.bbl}{}
 {\typeout{}
  \typeout{******************************************}
  \typeout{** Please run "bibtex \jobname" to optain}
  \typeout{** the bibliography and then re-run LaTeX}
  \typeout{** twice to fix the references!}
  \typeout{******************************************}
  \typeout{}
 }


\end{document}